\documentclass[prb,reprint,aps,amsmath,amssymb,showpacs]{revtex4-1}
\usepackage{graphicx}

\begin{document}
\preprint{}
\draft

\title{Extended linear regime of cavity-QED enhanced optical circular birefringence
induced by a charged quantum dot}

\author{C.Y.~Hu}\email{Chengyong.Hu@bristol.ac.uk}
\author{J.G.~Rarity} \email{John.Rarity@bristol.ac.uk}
\affiliation{Department of Electrical and Electronic Engineering, University of Bristol,
University Walk, Bristol BS8 1TR, United Kingdom}

\begin{abstract}

Giant optical Faraday rotation (GFR) and giant optical circular birefringence (GCB)
induced by a single quantum-dot spin in an optical microcavity can be regarded as
linear effects in the weak-excitation approximation if the input field lies in the
low-power limit [Hu et al, Phys.Rev. B {\bf 78}, 085307(2008) and ibid {\bf 80}, 205326(2009)].
In this work, we investigate the transition from the weak-excitation approximation
moving into the saturation regime comparing a semiclassical approximation with
the numerical results from a quantum optics toolbox [S.M. Tan, J. Opt. B {\bf 1},
424 (1999)]. We find that the GFR and GCB around the cavity resonance
in the strong coupling regime are input-field
independent at intermediate powers and can be well described by the semiclassical
approximation. Those associated with the dressed state resonances in the strong
coupling regime or merging with the cavity resonance in the Purcell regime are
sensitive to input field at intermediate powers, and cannot be well described
by the semiclassical approximation due to the quantum dot saturation. As the
GFR and GCB around the cavity resonance are relatively immune to the saturation
effects, the rapid read out of single electron spins can be carried out with
coherent state and other statistically fluctuating light fields. This also
shows that high speed quantum entangling gates, robust against input power
variations, can be built exploiting these linear effects.

\end{abstract}

\date{\today}

\pacs{78.67.Hc, 42.50.Pq, 78.20.Ek, 42.65.-k}

\maketitle

\section{Introduction}

Semiconductor charged quantum dots (QDs) with confined electron or hole spins
are promising for quantum computation, \cite{nielsen00, ladd10, liu10, awschalom13}
quantum communications,\cite{gisin02, scarani09, pan12} and quantum networks, \cite{cirac97}
especially for quantum internet with unconditional security. \cite{kimble08} Quantum gates are
the key components for quantum information processing in an analogue to the classical
gates for classical information processing. To design deterministic quantum gates,
three types of interactions can be exploited, i.e., photon-photon interactions, \cite{turchette95, imamoglu97, gheri98}
spin-spin interactions, \cite{loss98, imamoglu99, piermarocchi02, feng03, calarco03,clark07}
and photon-spin interactions. \cite{duan04, yao05, hu08, hu09, bonato10, hu11}
Although photons do not interact directly with each other intrinsically,
photon-photon indirect interactions mediated by cavity QED have been demonstrated
but are by definition non-linear phenomena. For high photon-photon gate fidelity it is thus
necessary to carefully control the \textquotedblleft shape\textquotedblright of the
overlapping photon wavepackets to be top hat profiles. Direct spin-spin interactions
suffer from short range distance. Among the three types of interactions, the
photon-spin interactions via optical transitions are the strongest and can be
easily configured to mediate photon-photon and spin-spin (indirect) interactions
for making various quantum gates with high speed.

Exploiting the cavity-QED enhanced photon-spin interactions, in our previous work
we proposed two types of photon-spin entangling gates consisting of a single
charged QD in an optical micro- or nano-cavity for both quantum and classical
information processing with high speed (tens to hundreds GHz) as well
as for spin memory with heralded feature and unity efficiency. \cite{hu08, hu09, hu11}
The two types of photon-spin entangling gates are based on the giant
optical Faraday rotation(GFR) and  giant optical circular birefringence(GCB), which
are induced by the single QD-confined spin in the cavity. GFR and GCB are manifested
as large differences in the phase or amplitude of reflection/transmission coefficients
between two circular polarizations of the input photons. Both phenomena can be
regarded as the macroscopic imprint of the optical spin selection rules of the charged
excitons in QDs.

However, in our previous work, the concepts of GFR and GCB were introduced in the
weak-excitation approximation where the input field is in the low-power limit, and
they can be regarded as the optical linear effects being independent of input power.
In this work, we investigate how GFR and GCB can be extended from the weak excitation
approximation to the semiclassical approximation where the QD saturation effects
induced by the input field are taken into account. An analytical method in the
semiclassical approximation is adopted in comparison with the numerical calculations
by the quantum optics toolbox. \cite{tan99}

We find that the semiclassical approximation can be used not only in the low- and high-
power regime, but also in a non-saturation window around the cavity resonance in the strong coupling
regime at intermediate  powers where the high cavity reflectivity leads to a higher
saturation threshold. This higher saturation threshold leads to the retention of the
linear effects into the intermediate power regime around the cavity resonance.
At frequencies close to the dressed state resonances, however, GFR and GCB become power
dependent at lower powers and saturates earlier. Similar low power saturation
occurs in the Purcell regime where there is no dressed state splitting. The quantum
gates based on the phase shifts(GFR) or reflection/transmission(GCB)
around the cavity resonance are thus much less vulnerable to input power fluctuations.

This work is organized as follows: In Sec. II, we work out an analytical
expression for the reflection coefficient in the semiclassical approximation
in the type-I spin-cavity system consisting of  a single QD spin in a
single-sided optical cavity. The reflection amplitude and phase spectra  are
calculated using both the analytical method and Tan's quantum optics toolbox.
The regions of linear and non-linear operation are identified and discussed.
In Sec. III, we derive the analytical expressions for the reflection and
transmission coefficients in the semiclassical approximation in the type-II
spin-cavity system with the single QD spin in a double-sided optical
cavity. The reflection and transmission spectra are calculated using both the
analytical method and Tan's quantum optics toolbox. We identify and analyze
the linear and nonlinear GCB. In Sec.IV, we show that the linear GFR and GCB
around the cavity mode resonance are not affected by the high-order dressed
state resonances.  In Sec.V, we summarize our conclusions.

\section{Linear and nonlinear GFR in type-I spin-cavity system}

A negatively (or positively) charged QD has an excess electron(or hole ) confined
in the QD. Charging a QD can be achieved by modulation doping techniques, or tunneling
in n-i-n structures.\cite{warburton13} The ground states of charged QD are two spin
states of the excess electron (or the excess hole), and the excited states are
two spin states of the negatively charged exciton $X^-$ (or the positively charged
exciton $X^+$) as shown in Fig. 1(c). Note that both the ground and the
excited states are spin degenerate due to the Kramer's theorem.\cite{exp0}

We consider such a charged QD embedded in a single-sided optical
microcavity or nanocavity with the one end mirror partially reflective
and another one $100\%$ reflective.\cite{hu08} The external light couples
the system via the partially reflective end mirror. Fig. 1(a) shows an experimental realization
with the pillar microcavity where two distributed Bragg reflectors (DBR) and transverse index
guiding provide three-dimensional confinement of light. The cross section of
the micropillar is made circular so that the cavity mode
are frequency degenerate for two circular polarizations. Some photonic crystal
nanocavities with specific symmetry (e.g., in Ref. \onlinecite{fushman08}) can
also support circularly polarized modes and are suitable for this work, too.
The cavity mode frequency is designed to match the optical transition
of QD.

\begin{figure}[ht]
\centering
\includegraphics* [bb= 104 152 525 750, clip, width=5cm, height=7.5cm]{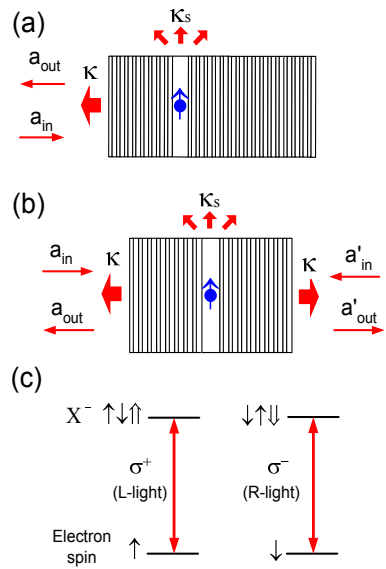}
\caption{(color online). (a) Type-I spin-cavity system consisting of a QD spin in
a single-sided cavity with one end mirror partially reflective and another end
mirror $100\%$ reflective. (b) Type-II spin-cavity system consisting of a QD
spin in a double-sided cavity with both end mirrors partially reflective. The
transmission of the two mirrors are made symmetric to achieve maximal resonant
transmission. (c) Optical spin selection rules for a negatively-charged
exciton $X^-$ in QD. Only the vertical transitions are shown here as the weak
cross transitions due to the heavy hole-light hole mixing can be corrected
and are thus neglected here.} \label{fig1}
\end{figure}

In this spin-cavity unit, there exists significant phase difference in the reflection
coefficients between the \textquotedblleft hot\textquotedblright and the
\textquotedblleft cold\textquotedblright cavity or between
two circular polarizations of the input photons. \cite{hu08} This GFR effect is a macroscopic
manifestation of the optical spin selection rule of charged excitons \cite{kheng93} [see
Fig. 1(c)] thanks to the cavity QED enhancement. The left circularly polarized photon (marked
as L or $\sigma^-$) only couples the transition $|\uparrow\rangle \leftrightarrow |\uparrow\downarrow\Uparrow\rangle$,
and the right circularly polarized photon (marked as R or $\sigma^+$) only couples
the transition $|\downarrow\rangle \leftrightarrow |\downarrow\uparrow\Downarrow\rangle$.
Here $|\uparrow\rangle$  and $|\downarrow\rangle$ represent electron spin states $|\pm
\frac{1}{2}\rangle$, $|\Uparrow\rangle$ and $|\Downarrow\rangle$ represent
heavy-hole spin states $|\pm\frac{3}{2}\rangle$ with the spin quantization axis
along the photon input direction. The photon polarizations are marked by the
input states to avoid any confusion due to the temporary polarization changes
upon reflection.

If the spin is in the state $|\uparrow\rangle$, the photon in the $|L\rangle$ state
couples to the cavity mode and feels like a \textquotedblleft hot\textquotedblright
cavity, whereas the photon in the $|R\rangle$ state does not couple to the cavity
mode and feels like a  \textquotedblleft cold\textquotedblright cavity. If the spin
is in the state $|\downarrow\rangle$, the photon in the $|R\rangle$ state feels like
a \textquotedblleft hot\textquotedblright cavity and the photon in the $|L\rangle$
state feels like a \textquotedblleft cold\textquotedblright cavity. The phase
difference of the reflection coefficient between the cold and hot cavity is mapped
to that between the two circular polarizations. Probing such a system with a linearly
polarized light leads to giant Faraday rotations of the polarization
directions of light (the GFR effect).\cite{hu08}

In the following, we extend the concept of GFR from the weak-excitation approximation
to the semiclassical approximation, and work out an analytical expression for the
reflection coefficient of the hot and cold cavity with the QD saturation effects taken
into account.

The Heisenberg equations of motions for the cavity field operator $\hat{a}$ and the QD dipole
operators $\sigma_-$, $\sigma_z$, \cite{walls94, kimble94} together with the input-output
relation \cite{gardiner85} can be written as
\begin{equation}
\begin{cases}
& \frac{d\hat{a}}{dt}=-\left[i(\omega_c-\omega)+\frac{\kappa}{2}+\frac{\kappa_s}{2}\right]\hat{a}-\text{g}\sigma_--\sqrt{\kappa}\hat{a}_{in} \\
& \frac{d\sigma_-}{dt}=-\left[i(\omega_{X^-}-\omega)+\frac{\gamma}{2}\right]\sigma_--\text{g}\sigma_z\hat{a}\\
& \frac{d\sigma_z}{dt}=2\text{g}(\sigma_+\hat{a}+\hat{a}^+\sigma_-)-\gamma_{\parallel}(1+\sigma_z)\\
& \hat{a}_{out}=\hat{a}_{in}+\sqrt{\kappa}\hat{a}
\end{cases}
\label{eqs1}
\end{equation}
where $\omega$, $\omega_c$, $\omega_{X^-}$ are the frequencies of the input field,
the cavity mode, and the $X^-$ transition, respectively.
g is the $X^-$-cavity coupling strength.\cite{exp2} $\gamma/2$ is the total QD dipole decay rate \cite{note1}
which includes the spontaneous emission induced decay rate $\gamma_{\parallel}/2$
and the pure dephasing rate $\gamma^*$, i.e., $\gamma/2=\gamma_{\parallel}/2+\gamma^*$.
$\kappa/2$ is the the cavity field decay rate into the input/output port. $\kappa_s/2$ is
the cavity field decay rate into the leaky modes due to side leakage, or other loss
channels such as the material background absorption and possible losses in the
highly-reflective end mirror in practical situation.

If the correlations between the cavity field and the QD dipole are neglected (this is
called the semiclassical approximation),\cite{allen87, armen87}
we have $\langle \sigma_{\pm}\hat{a}\rangle =\langle \sigma_{\pm}\rangle\langle \hat{a} \rangle$
and $\langle \sigma_z\hat{a}\rangle =\langle \sigma_z\rangle\langle \hat{a} \rangle$.
The conditions to apply the semiclassical approximation will be discussed later.
The reflection coefficient can thus be derived as
\begin{equation}
\begin{split}
r(\omega)& \equiv |r(\omega)|e^{i\phi(\omega)} \\
& =1-\frac{\kappa[i(\omega_{X^-}-\omega)+\frac{\gamma}{2}]}{[i(\omega_{X^-}-\omega)+
\frac{\gamma}{2}][i(\omega_c-\omega)+\frac{\kappa}{2}+\frac{\kappa_s}{2}]-\text{g}^2 \langle\sigma_z\rangle}.
\end{split}
\label{eqs2}
\end{equation}

The population difference  $\langle\sigma_z\rangle$ is given by
\begin{equation}
\langle\sigma_z\rangle=-\frac{1}{1+\frac{\langle n\rangle}{n_c[1+4(\omega_{X^-}-\omega)^2/\gamma^2]}},
\label{eqs3a}
\end{equation}
and the average cavity photon number $\langle n\rangle \equiv \langle \hat{a}^{\dagger}\hat{a}\rangle$ by
\begin{widetext}
\begin{equation}
 \langle n\rangle=\frac{\kappa[(\omega_{X^-}-\omega)^2+\frac{\gamma^2}{4}]P_{in}}
{[(\omega_{X^-}-\omega)^2+\frac{\gamma^2}{4}][(\omega_c-\omega)^2+\frac{(\kappa+\kappa_s)^2}{4}]+2\text{g}^2\langle\sigma_z\rangle
[(\omega_{X^-}-\omega)(\omega_c-\omega)-\frac{(\kappa+\kappa_s)\gamma}{4}]+\text{g}^4\langle\sigma_z\rangle^2},
\label{eqs3b}
\end{equation}
\end{widetext}
where $n_c=\gamma_{\parallel}\gamma/8\mathrm{g}^2$ is the critical photon number
which measures the average cavity photon number required to saturate the QD response,\cite{note2}
and $n_c=2.2 \times 10^{-4}$ is taken in this work.
$P_{in}$ is the input field power.
$\langle\sigma_z\rangle$ is the QD population difference between the excited state and the
ground state, and can be used to measure the saturation degree. $\langle\sigma_z\rangle$
ranges from $-1$ to $0$. If $\langle\sigma_z\rangle=-1$, QD is in the ground state (not saturated);
if $\langle\sigma_z\rangle=0$, QD is fully saturated, i.e., $50\%$ probability
in the ground states and $50\%$ probability in the excited states. If $\langle\sigma_z\rangle$
takes other values, the QD is partially saturated.

By solving Eqs. (\ref{eqs3a}) and (\ref{eqs3b}),  $\langle\sigma_z\rangle$ and $\langle n\rangle$
can be obtained at any input field strength. \cite{exp1} Note that $\langle\sigma_z\rangle$
and $\langle n\rangle$ are dependent on the input power, the frequency and the coupling
strength g. Putting $\langle\sigma_z\rangle$ into Eq. (\ref{eqs2}), we can obtain both the amplitude
and the phase of the reflection coefficient.

Alternatively, the reflection coefficient can be calculated numerically by the master equations in the
Lindblad form  with Tan's quantum optics toolbox.\cite{tan99}
The master equation for the spin-cavity system can be written as
\begin{equation}
\begin{split}
\frac{d\rho}{dt}= &-i[H_{JC},\rho]+(\kappa+\kappa_s)(\hat{a}\rho \hat{a}^{\dagger}-\frac{1}{2}\hat{a}^{\dagger}\hat{a}\rho-\frac{1}{2}\rho\hat{a}^{\dagger}\hat{a})\\
& +\gamma_{\parallel}(\hat{\sigma}_-\rho \hat{\sigma}_+ - \frac{1}{2}\hat{\sigma}_+\hat{\sigma}_-\rho-\frac{1}{2}\rho\hat{\sigma}_+\hat{\sigma}_-)+
\frac{\gamma^*}{2}(\hat{\sigma}_z\rho\hat{\sigma}_z-\rho)\\
\equiv & \mathcal{L}\rho,
\end{split}\label{master1}
\end{equation}
where $\rho$ is the reduced density matrix of the system, and all the parameters $\kappa, \kappa_s, \gamma, \gamma_{\parallel}, \gamma^{*}$ are defined
in the same way as in Eq.(\ref{eqs1}).  $\mathcal{L}$ is the Liouvillian
and $H_{JC}$ is the driven Jaynes - Cummings Hamiltonian with the input field driving the cavity.
In the frame
rotating at the input field frequency, $H_{JC}$ can be written as
\begin{equation}
\begin{split}
H_{JC}= &(\omega_c-\omega)\hat{a}^{\dagger}\hat{a}+(\omega_{X^-}-\omega)\sigma_+\sigma_-\\
& +ig(\sigma_+\hat{a}-\hat{a}^{\dagger}\sigma_-)+i\sqrt{\kappa}\hat{a}_{in}(\hat{a}-\hat{a}^{\dagger}),
\end{split}
\end{equation}
where  the input field is associated with the output field and the
cavity field by the input-output relation $\hat{a}_{out}=\hat{a}_{in}+\sqrt{\kappa}\hat{a}$ as
described earlier.

Although the analytical solution to the master equation in Eq. (\ref{master1}) is difficult,
Tan's quantum optics toolbox in Matlab provides an exact numerical solution to the density
matrix $\rho(t)$ or $\rho(t\rightarrow \infty)$ in steady state.
By taking the operator average in the input-output relation,
the reflection coefficient in the steady state can be calculated by the following expression
\begin{equation}
r(\omega)=1+\sqrt{\kappa}\frac{\mathrm{Tr}(\rho \hat{a})}{\langle\hat{a}_{in}\rangle},
\label{mtrx}
\end{equation}
This method yields the reflection coefficient for arbitrary input states
in principle. In this work we look at reflection coefficients for
classical input fields (coherent states) or single-photon trains
at different light intensities. The coherence time of these input fields are
long compared to the cavity lifetime.

\begin{figure}[ht]
\centering
\includegraphics* [bb= 65 415 491 772, clip, width=8cm, height=7cm]{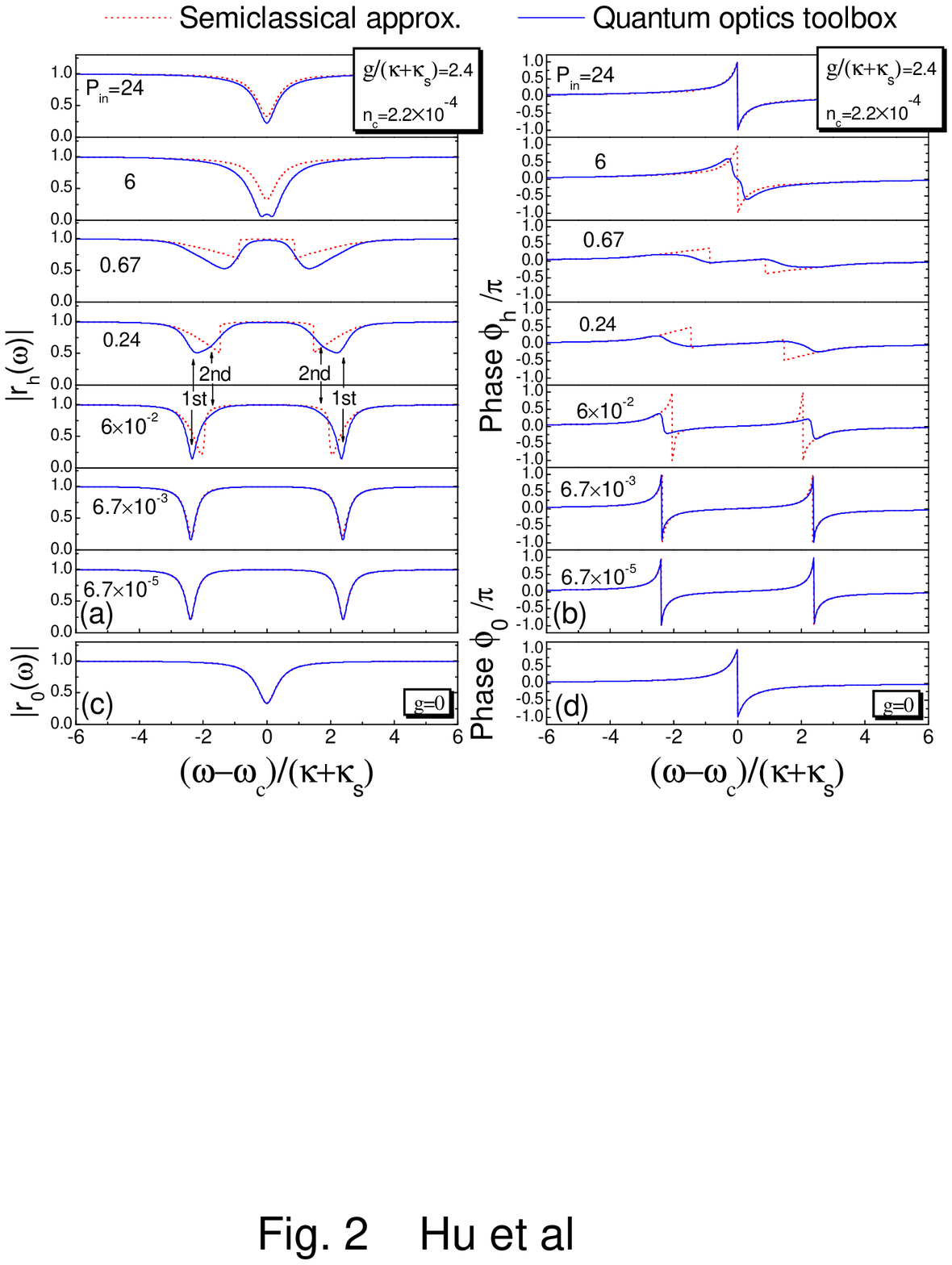}
\caption{(color online). (a) Reflectance $|r_h(\omega)|$ spectra and (b) phase
$\phi_h(\omega)$ spectra from a hot cavity with $\mathrm{g}=2.4(\kappa+\kappa_s)$
in the strong coupling regime at different input field powers. The input power $P_{in}$
is normalized by $\kappa+\kappa_s$ (i.e., in photons per cavity lifetime).
(c) Reflectance $|r_0(\omega)|$ spectra and (d) phase $\phi_0(\omega)$ spectra
from a cold cavity (with $\mathrm{g}=0$). Red dotted curves are calculated by
using Eq. (\ref{eqs2}) in semiclassical approximation, and blue solid curves
are calculated by the quantum optics toolbox. \cite{tan99}} \label{fig2}
\end{figure}

\begin{figure}[ht]
\centering
\includegraphics* [bb= 62 455 493 772, clip, width=8cm, height=7cm]{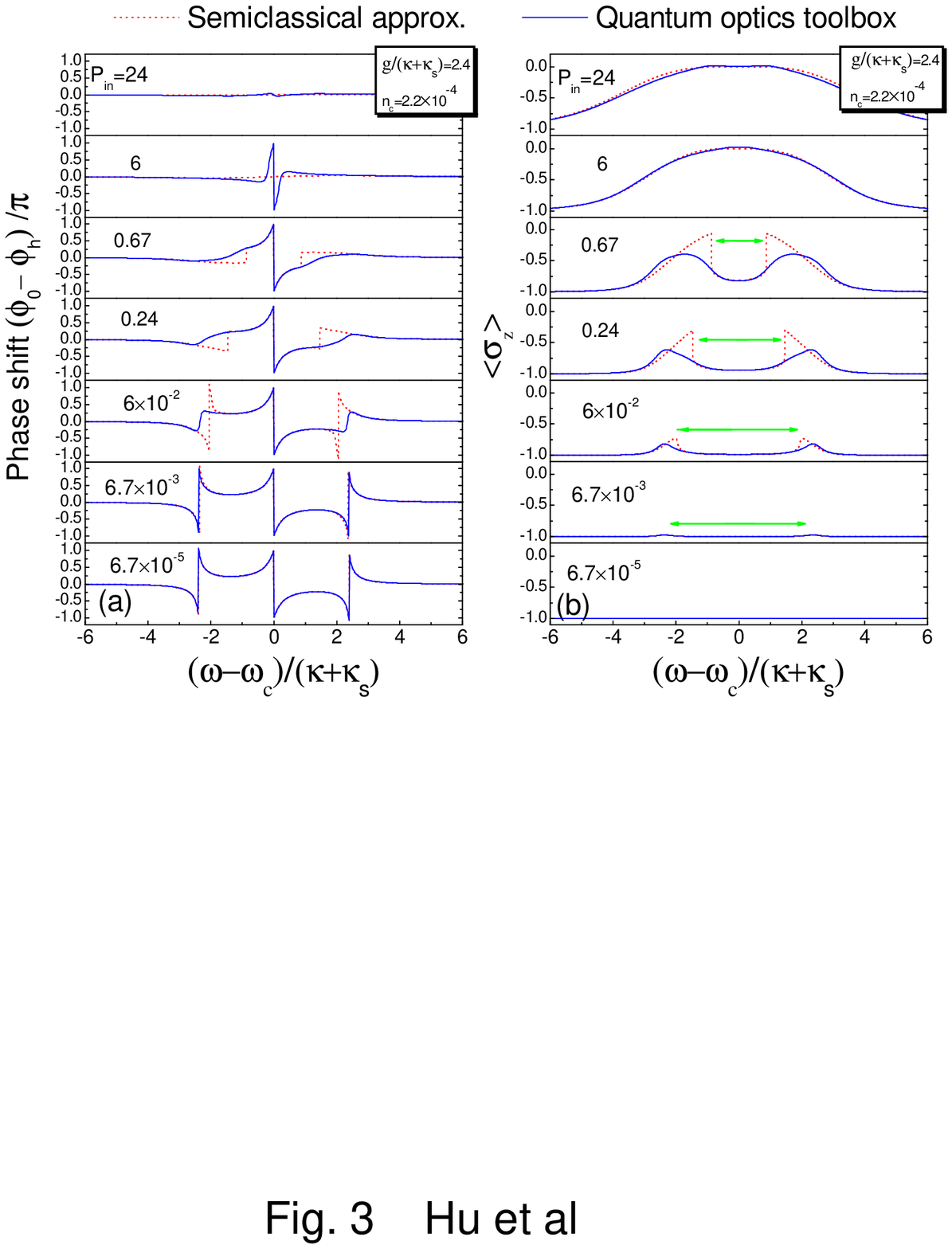}
\caption{(color online). (a) Phase shift $\phi_0(\omega)-\phi_h(\omega)$ spectra
between the cold cavity with $\mathrm{g}=0$ and the hot cavity with
$\mathrm{g}=2.4(\kappa+\kappa_s)$ at different input powers. The GFR angles
equal to one half of the phase shift.(b) The QD saturation curves at different
input powers. The non-saturation windows (marked by green arrows) are observed
between two dressed state resonances at intermediate powers. The input power is normalized
by $\kappa+\kappa_s$ (i.e., in photons per cavity lifetime). Red dotted curves are
calculated by using Eq. (\ref{eqs2}) in semiclassical approximation, and blue solid
curves are calculated by the quantum optics toolbox.} \label{fig3}
\end{figure}

Next we study the reflection spectra calculated from the two methods described
above. We focus on the results in the strong coupling regime as from these
results the information in the Purcell regime or weak coupling regime can be extracted.
Strongly coupled QD-cavity systems with $\mathrm{g}>(\kappa+\kappa_s-\gamma)/4$ have been
experimentally demonstrated in various micro- or nano-cavities. \cite{reithmaier04, yoshie04, peter05}
In this work we take $\mathrm{g}=2.4(\kappa+\kappa_s)$ which can be achieved
for In(Ga)As QDs in the state-of-the-art pillar microcavity.\cite{reithmaier04, reitzenstein07, volz12}
The side leakage rate $\kappa_s$ depends on fabrication and various cavity details such as
materials, structures, size, etc.,  and we take $\kappa_s=0.5\kappa$ in our calculations.
The total QD decay rate $\gamma$ due to the spontaneous emission and the pure
dephasing processes is sample dependent and it is usually smaller than
the cavity decay rate $\kappa+\kappa_s$ in high-quality samples.
The QD is tuned in resonance with the cavity mode, i.e., $\omega_{X^-}=\omega_c=\omega_0$.

Fig. 2(a) and 2(b) show the reflectance $|r(\omega)|$ and phase $\phi(\omega)$
spectra of the hot cavity in the strong coupling regime with $\mathrm{g}/(\kappa+\kappa_s)=2.4$
at different input powers. The input power is normalized by $(\kappa+\kappa_s)$ (i.e.,
in photons per cavity lifetime) where $\kappa+\kappa_s$ is the total cavity decay rate.
At low powers ($P_{in}<0.026$ photons/cavity lifetime, low power regime), the two dips
observed in the reflectance spectra and the related two oscillating features in the phase
spectra due to the resonances of  the first manifold of  dressed states (also called polariton states or normal modes)
separated by the vacuum Rabi splitting (or normal mode splitting) \cite{note3}.
We note that the semiclassical approximation and the toolbox yield identical results.

With increasing input power ($0.026<P_{in}<1.1$, intermediate power regime), the
two reflectance dips and phase features become weaker, and both shift towards the
cavity resonance at $\omega_c$ (i.e., at the zero detuning $\omega-\omega_c=0$).
Moreover, there are obvious discrepancy on the reflectance dips and phase features
between the two calculation methods. Further increasing the power ($P_{in} > 1.1$, high power regime),
the two reflectance dips and phase features merge into one around the cavity resonance.
Both the reflectance and the phase spectra of the hot cavity look similar to those of
the cold cavity as shown in Fig. 2(c) and Fig. 2(d). We note that the power-dependent
reflection spectra \cite{exp3} were experimentally demonstrated recently. \cite{loo12, englund12, bose12}

The phase difference between the cold and hot cavity is  an indication of the phase
difference between two circular polarizations of the reflected photons if the spin
is included (see discussions at the beginning of this section), which is the GFR effect.
Fig.3(a) present the phase difference $\phi_0(\omega)-\phi_h(\omega)$  spectra between
the cold and the hot cavity at different input powers.  Note that the GFR angle equals
to one half of the phase difference. Besides the two oscillating phase features associated
with the dressed state resonances, there is another oscillating feature around the cavity
resonance. The third phase feature is mainly contributed by the cold cavity as the phase
is nearly zero around the cavity frequency for the hot cavity as shown in Fig. 2(b).
The strength of the third phase feature is not affected by the input field in the
low and intermediate power regime, but it disappears in the high power regime. We also note
that the semiclassical approximation works well for this phase feature as both calculations
yield the same results. The other two phase features related to the dressed states
shift toward the cavity resonance, merge into one and finally disappear with increasing
the input power in the intermediate and high power regime. For these two phase
features, there are significant discrepancies between the semiclassical approximation
and the toolbox.

The above results can be explained by the QD saturation induced by the input field.
The saturation spectra is shown in Fig. 3(b). In the low power regime ($P_{in}\ll1$), the saturation
effect can be neglected as the QD is almost in the ground state, i.e.,  $\langle \sigma_z\rangle \simeq -1$
in the whole frequency range. This is exactly the weak-excitation approximation used
in our previous work.\cite{hu08, hu09} As there is no real excitation, there is no
correlations between the cavity field and the QD dipole, and the assumption $\langle \sigma_{\pm,z}\hat{a}\rangle
=\langle \sigma_{\pm,z}\rangle\langle \hat{a} \rangle$ is valid. The semiclassical
approximation is then equivalent to the the weak-excitation approximation in the low power
regime.  This explains why the reflectance, the phase and the phase shift or GFR spectra
are not affected by input field at low  powers.

In the intermediate power regime ($P_{in}\sim1$), the input field in resonance with  the dressed states can
enter the cavity and build the cavity field which saturates the QD. The QD saturation reduces
the QD-cavity coupling strength to $\mathrm{g}_{eff}=\mathrm{g}\sqrt{|\langle \sigma_z\rangle|}$,
so the Rabi splitting becomes smaller and the dressed state resonances shift towards the cavity
resonance frequency with increasing input powers ($\langle\sigma_z\rangle$ from $-1$ to $0$)[see
Figs. 2 and 3]. The phase difference (or GFR) associated with the dressed state resonance is nonlinear
as both the strength and the frequency vary with the input field. There are
significant correlations between the QD dipole and the cavity fields, so the semiclassical
approximation does not work well for this nonlinear GFR associated with the dressed
state resonances.

\begin{figure}[ht]
\centering
\includegraphics* [bb= 108 461 452 752, clip, width=6cm, height=6cm]{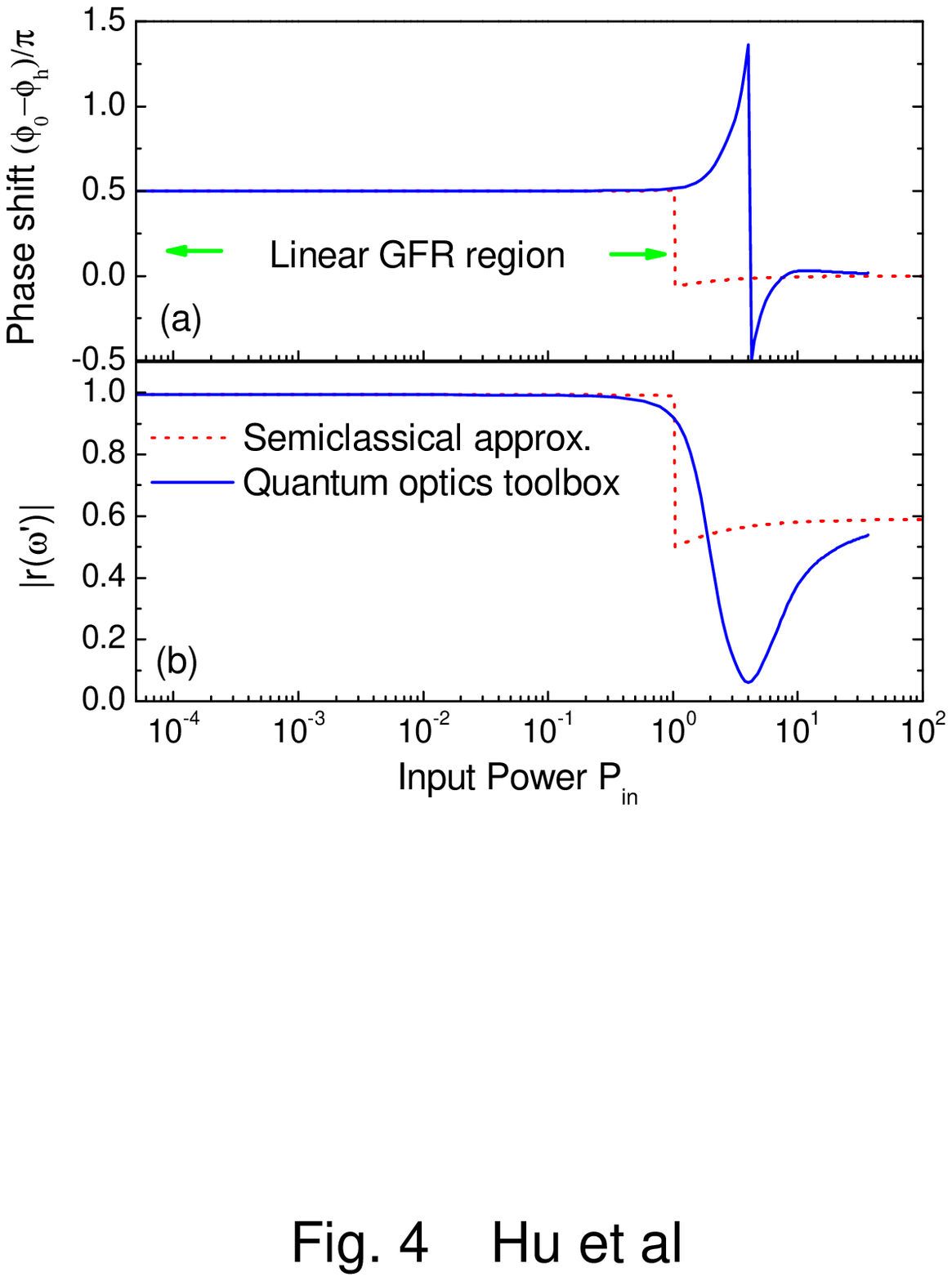}
\caption{(color online). (a) The linear phase shift (corresponding to the linear GFR) as
a function of the input power. The phase shift is tracked at a fixed frequency within
the non-saturation window. Here we choose a phase shift of $\pi/2$, corresponding to a
GFR angle of $\pi/4$. (b) The reflectance as a function of the input power. The
reflectance is tracked at a fixed frequency which is the same as in (a). The input power
is normalized by $\kappa+\kappa_s$ (i.e., in photons per cavity lifetime). Red dotted
curves show results calculated from the semiclassical approximation, and blue solid curves
are calculated by the quantum optics toolbox.} \label{fig4}
\end{figure}

However,in the intermediate power regime the saturation effect remains weak around
$\omega=\omega_c$ [see Fig. 3(b)]. The cavity resonance is a highly reflective
region [see Fig. 2(a)] which prevents photons from entering the cavity and saturating
the QD effectively. Therefore, the QD remains in the ground states,
i.e., $\langle \sigma_z\rangle \simeq -1$. In this non-saturation window, the semiclassical
approximation still works well, and yields the same results as the toolbox. The GFR spectra
in the non-saturation window is not affected by the input power, and remain
the same strength as that in the low power limit except the window size
shrinks with increasing input power. The GFR within the non-saturation window
is therefore a linear effect.

In the high power regime($P_{in}\gg1$), the saturation effect becomes so strong that the non-saturation
window is closed,\cite{exp4} and the QD is fully saturated, i.e., $\langle \sigma_z\rangle \simeq 0$.
The full saturation starts from the center of the cavity resonance and extends
towards its two sides. The cavity with a saturated QD behaves like a cold
cavity, so the phase difference between the hot and cold cavity
disappears [see Fig. 3(a)].

Besides the evolution of the non-saturation window, from the saturation spectra
we also observe the power or saturation broadening of the QD response
with increasing the input power. \cite{loudon03}

Fig.4 (a) presents GFR at a fixed frequency close to the cavity resonance (within the
non-saturation window) as a function of input power. The phase difference of $\pi/2$ is chosen
as the  $\pi/2$ phase shift is required for making
the ideal photon-spin entangling gates. \cite{hu08} We see that GFR is constant with
the input power up to $P_{in} \simeq 1$ in accordance with the calculations. \cite{exp4}
 In this region,
the reflectance is also independent of the input power as shown in Fig. 4(b). As a result,
the photon-spin entangling gate (see ref.\onlinecite{hu08}) based on this linear GFR is
resistant to the photon rate variations, which is highly desirable in practical applications such as
quantum communications and quantum computation. Again it is verified that the semiclassical
approximation works well for the linear GFR as it yields identical results as the
toolbox. In contrast, the GFR related to the dressed state resonances
start saturating at much lower powers [see Fig. 3(a)].

The large fluctuations of GFR at the start of high power regime (see Fig. 4) are due to the
nonlinear GFR. When the input power increases from the intermediate to the high
power regime, the saturation induces a transition from the strong coupling
to weak coupling regime at an input power where the non-saturation window is closed
and the linear GFR was eaten up by the nonlinear GFR. This also
takes place in the conventional Purcell regime with
$\gamma < 4\text{g}^2/(\kappa+\kappa_s)<(\kappa+\kappa_s)$ where no linear GFR exists
and the cavity resonance region is covered by the nonlinear GFR that is vulnerable
to the input field power (the results are not shown here). When the QD saturation
occurs over a large frequency range, the nonlinear GFR disappears as well. Similarly, in the
weak coupling regime $4\mathrm{g}^2/(\kappa+\kappa_s)<\gamma$, the concept of
\textquotedblleft one-dimensional atom\textquotedblright breaks down and no GFR
exists. Instead, the conventional Faraday rotation enhanced by the cavity  (due to
the back and forth propagation of light in the cavity)  can be observed with
the rotation angles at least five orders of magnitude smaller than the GFR angles.
Note that GFR is enhanced by the cavity QED, \cite{hu08} rather than the cavity only.

\section{Linear and nonlinear GCB in type-II spin-cavity system}

In this Section, we consider the type-II spin-cavity unit with a charged QD in a
double-sided optical microcavity where the two end mirrors are both partially
reflective [see Fig. 1(b)]. In this spin cavity-QED system, the GCB manifests
as the different reflection/transmission coefficients between the cold and hot
cavity or between the R- and L- circular polarizations of input photons. This
allows us to make another photon-spin entangling gate, i.e., the entanglement
beam splitter,\cite{hu09} which  can directly split a spin-photon polarization
product state into two constituent entangled states.

The reflection or transmission behaviors in similar systems were investigated
in the weak coupling regime in either the weak-excitation approximation \cite{shen05, waks06, garnier07}
or in the semiclassical approximation. \cite{garnier07, majumdar12} To derive the
reflection and transmission coefficients in the strong coupling regime with the QD
saturation taken into account, we apply the same approach as discussed in
Sec.II , i.e., the Heisenberg equations of motion for the cavity field
operator $\hat{a}$ and the QD dipole operators $\sigma_-$ and
$\sigma_z$, \cite{walls94, kimble94} together with the input-output
relations, \cite{gardiner85}
\begin{equation}
\begin{cases}
& \frac{d\hat{a}}{dt}=-\left[i(\omega_c-\omega)+\kappa+\frac{\kappa_s}{2}\right]\hat{a}-\text{g}\sigma_- \\
& ~~~~~~ -\sqrt{\kappa}\hat{a}_{in}-\sqrt{\kappa}\hat{a}^{\prime}_{in}\\
& \frac{d\sigma_-}{dt}=-\left[i(\omega_{X^-}-\omega)+\frac{\gamma}{2}\right]\sigma_--\text{g}\sigma_z\hat{a} \\
& \frac{d\sigma_z}{dt}=2\text{g}(\sigma_+\hat{a}+\hat{a}^+\sigma_-)-\gamma_{\parallel}(1+\sigma_z) \\
& \hat{a}_{out}=\hat{a}_{in}+\sqrt{\kappa}\hat{a} \\
& \hat{a}^{\prime}_{out}=\hat{a}^{\prime}_{in}+\sqrt{\kappa}\hat{a}. \\
\end{cases}
\label{eqd1}
\end{equation}
All the parameters here have the same definitions and meanings as in Eq. (\ref{eqs1}).

Following a similar procedure in Sec. II, the analytical expressions for the reflection
and transmission coefficients can be derived in the semiclassical approximation
by neglecting the correlations between the cavity field and the QD dipole,i.e.,
\begin{equation}
\begin{split}
& r(\omega)=1+t(\omega), \\
& t(\omega)=\frac{-\kappa[i(\omega_{X^-}-\omega)+\frac{\gamma}{2}]}{[i(\omega_{X^-}-\omega)+
\frac{\gamma}{2}][i(\omega_c-\omega)+\kappa+\frac{\kappa_s}{2}]-\text{g}^2 \langle \sigma_z\rangle}.
\end{split}
\label{eqd2}
\end{equation}
The semiclassical approximation works well in three situations: (1) low-power
limit in the weak or strong coupling regime; (2) high-power limit in the weak or strong coupling
regime; (3) within the non-saturation window in the
strong coupling regime. These conditions to apply the semiclassical approximation are the same
for both types of spin-cavity systems.

The average population $\langle\sigma_z\rangle$ is given by Eq. (\ref{eqs3a}) and the
average cavity photon number $\langle n\rangle$ by
\begin{widetext}
\begin{equation}
\langle n\rangle=\frac{\kappa[(\omega_{X^-}-\omega)^2+\frac{\gamma^2}{4}]P_{in}}
{[(\omega_{X^-}-\omega)^2+\frac{\gamma^2}{4}][(\omega_c-\omega)^2+\frac{(2\kappa+\kappa_s)^2}{4}]+2\text{g}^2\langle\sigma_z\rangle
[(\omega_{X^-}-\omega)(\omega_c-\omega)-\frac{(2\kappa+\kappa_s)\gamma}{4}]+\text{g}^4\langle\sigma_z\rangle^2},
\label{eqd3}
\end{equation}
\end{widetext}
where the critical photon number $n_c$ is defined in the same way as in Sec. II.

From Eqs. (\ref{eqs3a}) and (\ref{eqd3}), both $\langle\sigma_z\rangle$ and
$\langle n\rangle$ can be calculated at any given input field strength. Putting
$\langle\sigma_z\rangle$  into Eq. (\ref{eqd2}), we can obtain the reflection and
transmission coefficients.
Alternatively, the reflection
and transmission coefficients can  be
calculated  numerically in the frame of master equation using Tan's quantum optics
toolbox with the same technique as described in Sec.II.
From the obtained density matrix $\rho$ in steady state and the input-output relations, the reflection
and transmission coefficients can be calculated by the following expressions
\begin{equation}
\begin{split}
& r(\omega)=1+t(\omega), \\
& t(\omega)=\sqrt{\kappa}\frac{\mathrm{Tr}(\rho \hat{a})}{\langle\hat{a}_{in}\rangle}.
\end{split}
\end{equation}
We use the above two methods to study the GCB in this spin-cavity system.

\begin{figure}[ht]
\centering
\includegraphics* [bb= 65 414 489 772, clip, width=8cm, height=7cm]{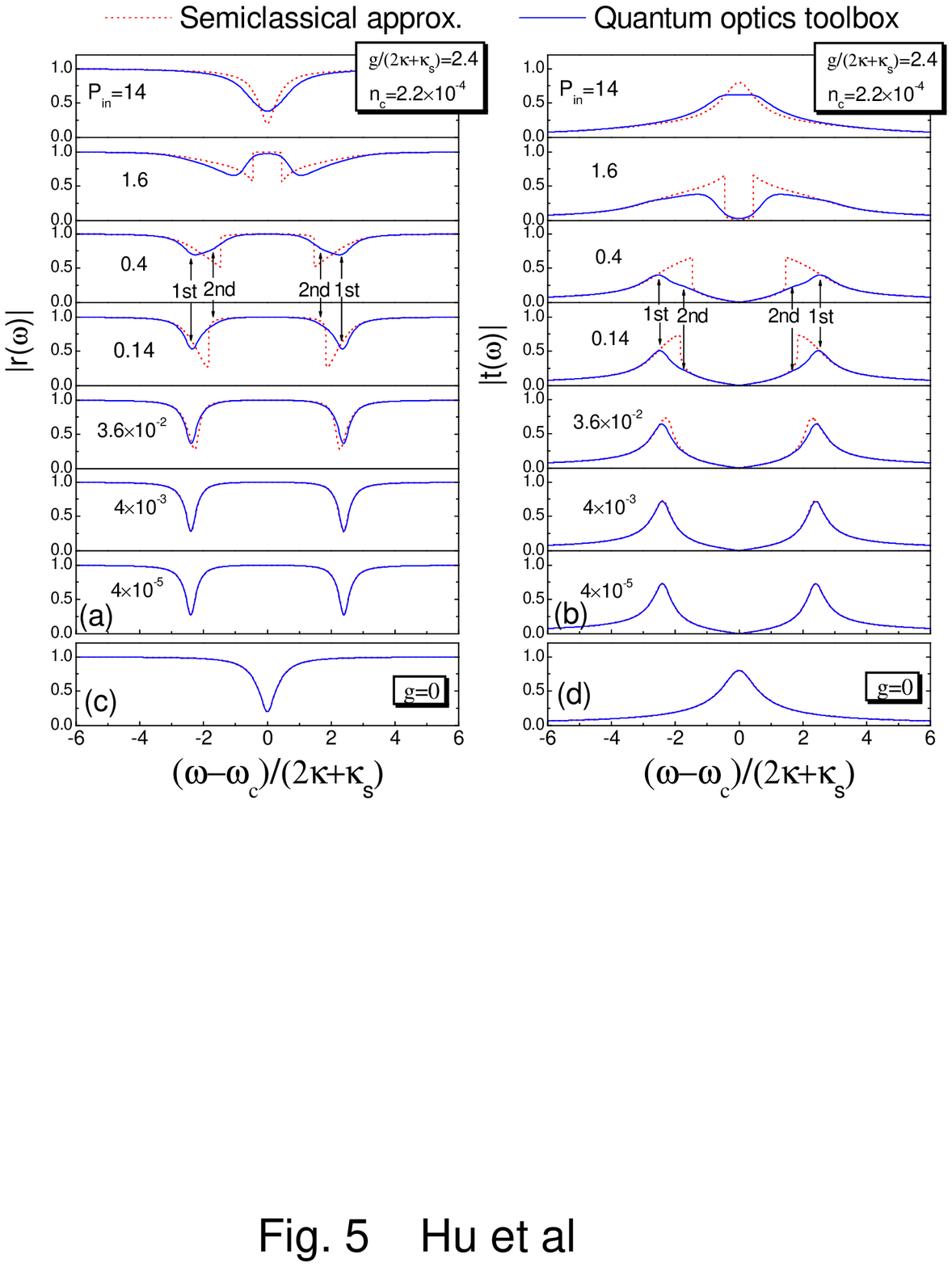}
\caption{(color online). (a) Reflectance $|r_h(\omega)|$ spectra and (b) transmittance
$|t_h(\omega)|$ spectra from a hot cavity in the strong coupling regime with
$\mathrm{g}=2.4(2\kappa+\kappa_s)$ at different input field powers.
(c) Reflectance $|r_0(\omega)|$ spectra and (d) transmittance  $|t_0(\omega)|$ spectra
from a cold cavity ($\mathrm{g}=0$). The input powers are normalized
by $2\kappa+\kappa_s$ (i.e., in photons per cavity lifetime). Red dotted curves
are calculated by using Eq. (\ref{eqd2}) in semiclassical approximation, and
blue solid curves are calculated by the quantum optics toolbox.} \label{fig5}
\end{figure}

Fig. 5(a) and 5(b) show the reflectance $|r(\omega)|$ and the transmittance $|t(\omega)|$
spectra of the hot cavity in the strong coupling regime with $\mathrm{g}/(2\kappa+\kappa_s)=2.4$
at different input powers. The reflection or transmission coefficients are
different between the hot [see Fig . 5(a) and 5(b)] and the cold  cavity [see Fig . 5(c) and 5(d)],
indicating the reflection or transmission difference between the two circular
polarizations of the input photons, which is  the GCB effect.\cite{hu09}
Note that the reflection or transmission coefficients of the cold cavity [see Fig . 5(c) and 5(d)]
are input-power independent, so the reflection or transmission coefficients of the hot cavity
can stand alone to represent the GCB effect [see Fig. 5(a) and (b)].

Following the similar discussions in Sec. II, we identify both linear and
nonlinear GCB in this spin-cavity unit. The linear GCB lie within the
non-saturation window around the cavity resonance and is manifested as nearly unity
reflectance and nearly zero transmission from the hot cavity. The linear GCB does
not depend on the input power in the low ($P_{in}<0.015$) and intermediate power
regime $0.015<P_{in}<1.7$ as the QD saturation within the non-saturation window
is negligibly small. However, the linear GCB disappears at the start of the high power
regime $P_{in}>1.7$ where the non-saturation window is closed.\cite{exp4}

The nonlinear GCB is associated with the dressed state resonances separated by the
vacuum Rabi splitting. It is manifested as the two dips in the reflection spectra
and two peaks in the transmission spectra. With increasing input power, the QD
saturation becomes significant, so the two reflection dips and the two transmission
peaks weaken and shift towards the cavity resonance. When the two reflection dips and
the two transmission peaks merge into one dip or
peak, the QD is fully saturated and the hot cavity turns to a cold
cavity. As a result, the nonlinear GCB disappears. We see that the input field induces
a transition from the strong coupling to Purcell regime and finally to the weak coupling
regime. This transition also happens in the type-I spin-cavity unit as discussed
in Sec. II.

\begin{figure}[ht]
\centering
\includegraphics* [bb= 107 461 452 752, clip, width=6cm, height=6cm]{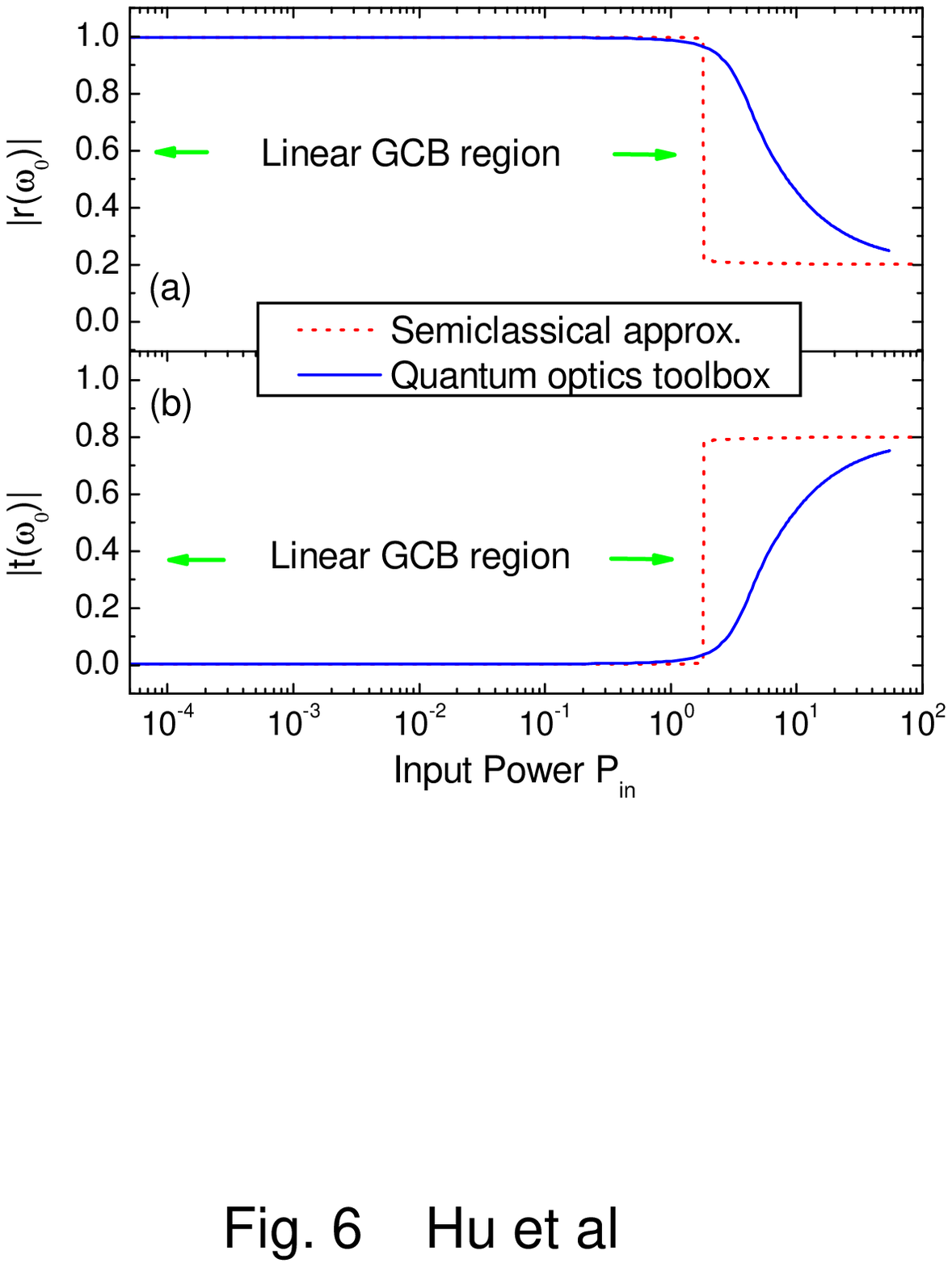}
\caption{(color online). Reflectance and transmittance at $\omega=\omega_c$ from the hot cavity
as a function of the input power. Note that the reflectance or transmittance from the cold
cavity are independent of input power (not shown here). The input powers are normalized by
$2\kappa+\kappa_s$ (i.e., in photons per cavity lifetime). Red dotted curves show results
from the semiclassical approximation, and blue solid curves are calculated by the
quantum optics toolbox.} \label{fig6}
\end{figure}

The linear GCB as a function of the input power is presented in Fig. 6 (a) and 6(b).
In the low and intermediate power regime, the reflectance $|r(\omega_0)|$
and the transmittance $|t(\omega_0)|$ at the center of the cavity resonance remain constant
with increasing input power up to $P_{in}\simeq 1.7$ in accordance with the calculations.\cite{exp4}
Similar to the linear GFR discussed in Sec. II, the quantum gates built from the
linear GCB are robust against the variations of input power or input photon rate.

In Purcell regime ($\gamma<4\mathrm{g}^2/(2\kappa+\kappa_s)<2\kappa+\kappa_s$),
the cavity resonance region is covered by the nonlinear GCB and there exists no
linear GCB. As the nonlinear GCB is input-power dependent, the quantum gates based on the
nonlinear GCB are fragile when the input power varies. In the weak coupling
regime $4\text{g}^2/(2\kappa+\kappa_s)<\gamma$ where  the concept of
\textquotedblleft one-dimensional atom\textquotedblright (Ref.\onlinecite{kimble94})
breaks down, there is no GCB effect. It is interesting to note that the concept
of \textquotedblleft one-dimensional atom\textquotedblright can be partially recovered
by placing QD or atom in one-dimensional waveguides. However, the GCB expected in this
waveguide structures is a nonlinear effect, therefore it is sensitive to the QD saturation
or the input power.

\section{Influence of high-order dressed states on linear GFR and GCB}

All results presented in previous sections are calculated in the strong coupling regime with
$\mathrm{g}/(\kappa+\kappa_s)=2.4$ for single-sided cavity [or
$\mathrm{g}/(2\kappa+\kappa_s)=2.4$ for double-sided cavity] which can be experimentally achieved.
The two reflection dips or two transmission peaks are explained as the first
order dressed state resonances,i.e., the  transitions $|0, G\rangle \rightarrow |1, \pm\rangle$
where $|0, G\rangle$ is the ground state, and $|1, \pm\rangle$ is the first-order
dressed states with $ n=1$ excitation.\cite{note3}  In the strong coupling regime, we also expect
high-order dressed state resonances $|0, G\rangle \rightarrow |n, \pm\rangle$(see
Fig. 7). These dressed states build the anharmonic energy level diagram described by the
Jaynes - Cummings ladder \cite{jaynes63} which is regarded as an indication of quantum nature
of light-matter interactions. In Figs. (2) and (5) only very weak resonances
related to the $n=2$ dressed states are identified in the reflection and transmission
spectra when the input power is around $P_{in}=0.1-0.4$, and resonance peaks related to high-order dressed
states ($n>2$) are washed out by the QD saturation, the cavity photon probability
distribution and the resonance broadening.

\begin{figure}[ht]
\centering
\includegraphics* [bb= 69 272 561 632, clip, width=8cm, height=6.4cm]{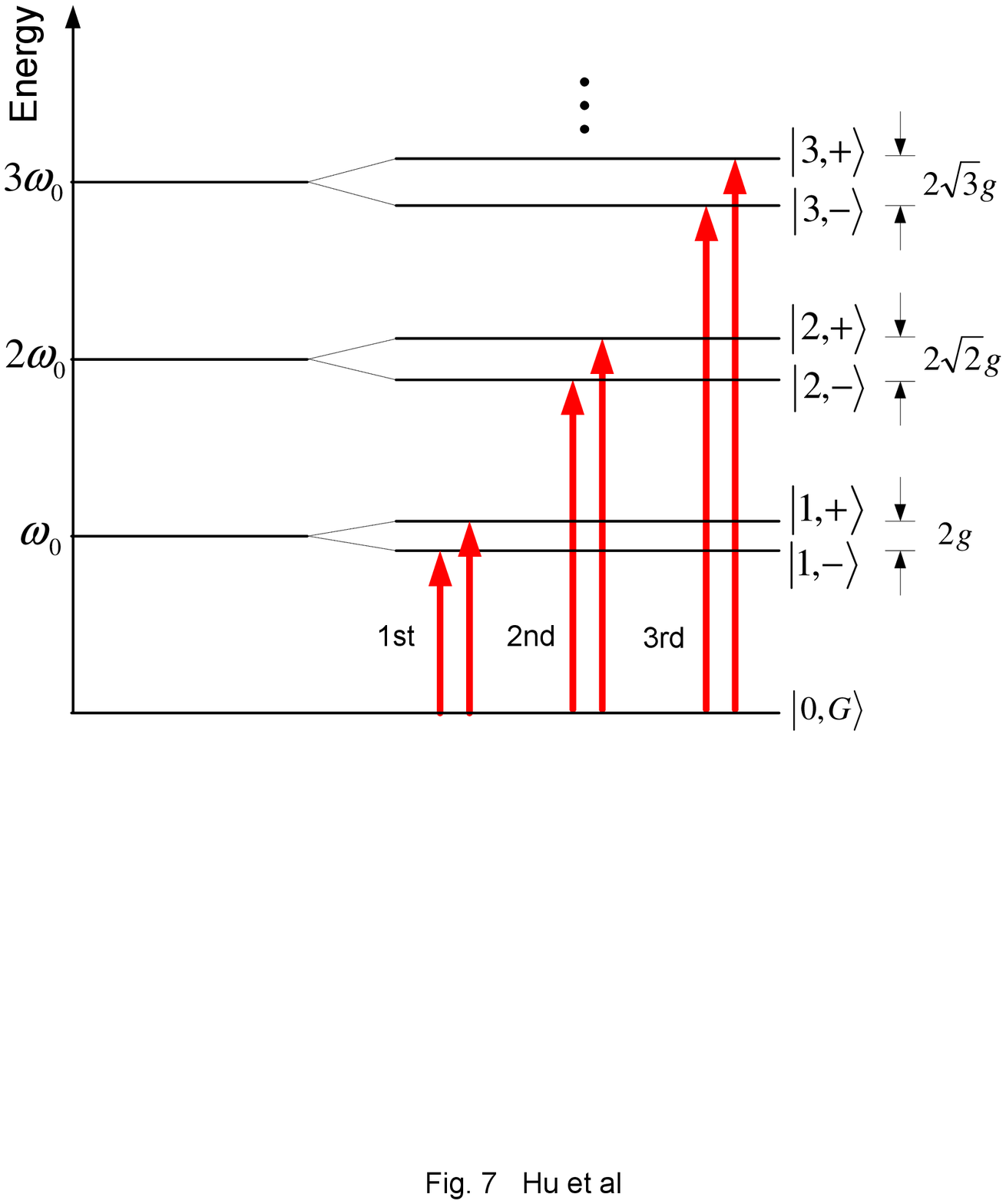}
\caption{(color online). The Jaynes - Cummings energy spectrum and n-photon transitions
from the ground state $|0, G\rangle$ to the dressed states $|n, \pm\rangle$. These high-order,
multi-photon transitions can be observed in reflection or transmission spectra
under some conditions as discussed in text. Dissipation processes are neglected
in this diagram. } \label{fig7}
\end{figure}

From Eqs. (\ref{eqs3a}) and (\ref{eqs3b}), we see that around
the dressed state resonances or the cavity resonance the QD becomes less saturated and there are less
photons accumulated in the cavity with increasing the coupling strength (if the
input power is kept the same). At higher coupling strength,
we clearly observe the high-order dressed state resonances around
the two edges of the non-saturation window. The larger the coupling strength,
the more dressed state resonances can be observed. Although these large coupling
strength $\mathrm{g}/(\kappa+\kappa_s)$ for QD-cavity systems go beyond the current
state-of-the-art value $\mathrm{g}/(\kappa+\kappa_s)=2.7$,\cite{volz12} this deeper
strong coupling regime allows us to investigate
whether or not these high-order dressed states  affect the linearity of GFR and GCB,
and meanwhile get more insight into the spin-cavity QED systems.

\begin{figure}[ht]
\centering
\includegraphics* [bb= 63 452 497 779, clip, width=8cm, height=7cm]{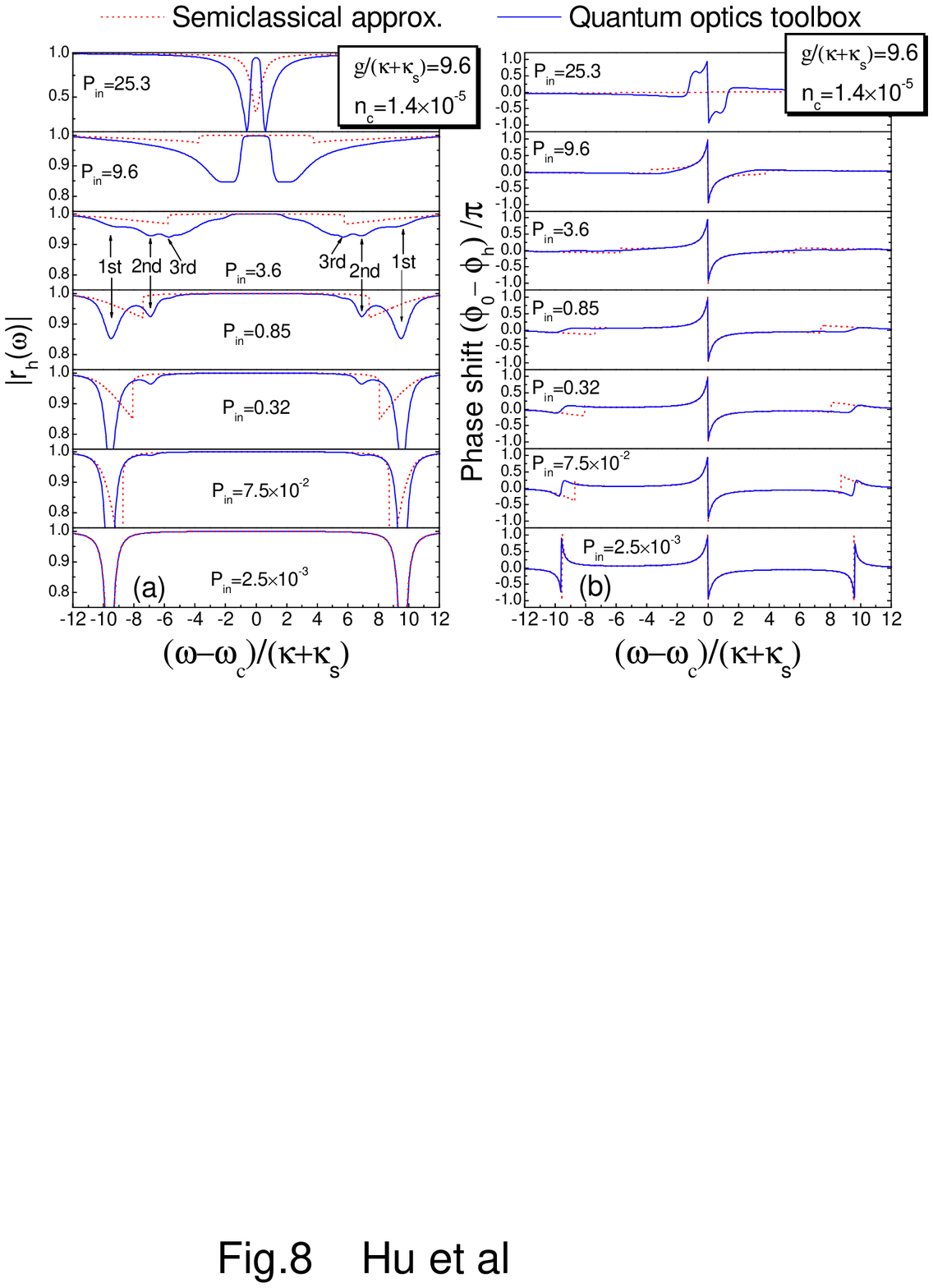}
\caption{(color online). (a) Reflectance $|r_h(\omega)|$ spectra and (b) phase
$\phi_h(\omega)$ spectra from a hot cavity with $\mathrm{g}=9.6(\kappa+\kappa_s)$
in the strong coupling regime at different input field powers. The input power $P_{in}$
is normalized by $\kappa+\kappa_s$ (i.e., in photons per cavity lifetime).
The 1st, 2nd and 3rd manifold of dressed states are observed in certain power range.
Red dotted curves are calculated by
using Eq. (\ref{eqs2}) in semiclassical approximation, and blue solid curves
are calculated by the quantum optics toolbox.} \label{fig8}
\end{figure}

Fig.8 presents the reflection and phase shift spectra at different input powers for a
single-sided spin-cavity system with $\mathrm{g}/(\kappa+\kappa_s)=9.6$. The $n=1,2,3$ dressed
states are identified in reflection spectra calculated by the toolbox, but not by the semi-classical model.
The first manifold
of dressed states are observed for input powers below $P_{in}=9$, and the second
manifold for input powers between $P_{in}=0.07$ and $P_{in}=9$, and the third manifold between
$P_{in}=1.3$ and $P_{in}=9$. The dressed state resonances shift towards
the cavity resonance with increasing the order n, and satisfy the resonance condition
$n\omega=\omega_{n,\pm}$
where $\omega$ is the frequency of input field and $\omega_{n,\pm}$ are the
energy eigenvalues of the nth-order dressed states. By diagonalization of
the Liouvillian defined in Eq.(\ref{master1}), $\omega_{n,\pm}$ can be derived as
$\omega_{n,\pm}= n\omega_0 -i[(2n-1)(\kappa+\kappa_s)+\gamma]/4
\pm\sqrt{n\mathrm{g}^2-[(\kappa+\kappa_s-\gamma)/4]^2}$
under the weak-excitation condition at zero detuning.
As these dressed states $|n, \pm\rangle \simeq (|n, G\rangle \pm |n-1, E\rangle)/\sqrt{2}$
are highly entangled states between QD and cavity field, it is not surprising that the corresponding dressed
state resonances cannot be observed in the reflection spectra calculated by the semi-classical model
which neglects the correlation between QD and the cavity field [see Figs.(\ref{fig8})].\cite{note3}

\begin{figure}[ht]
\centering
\includegraphics* [bb= 62 458 488 775, clip, width=8cm, height=7cm]{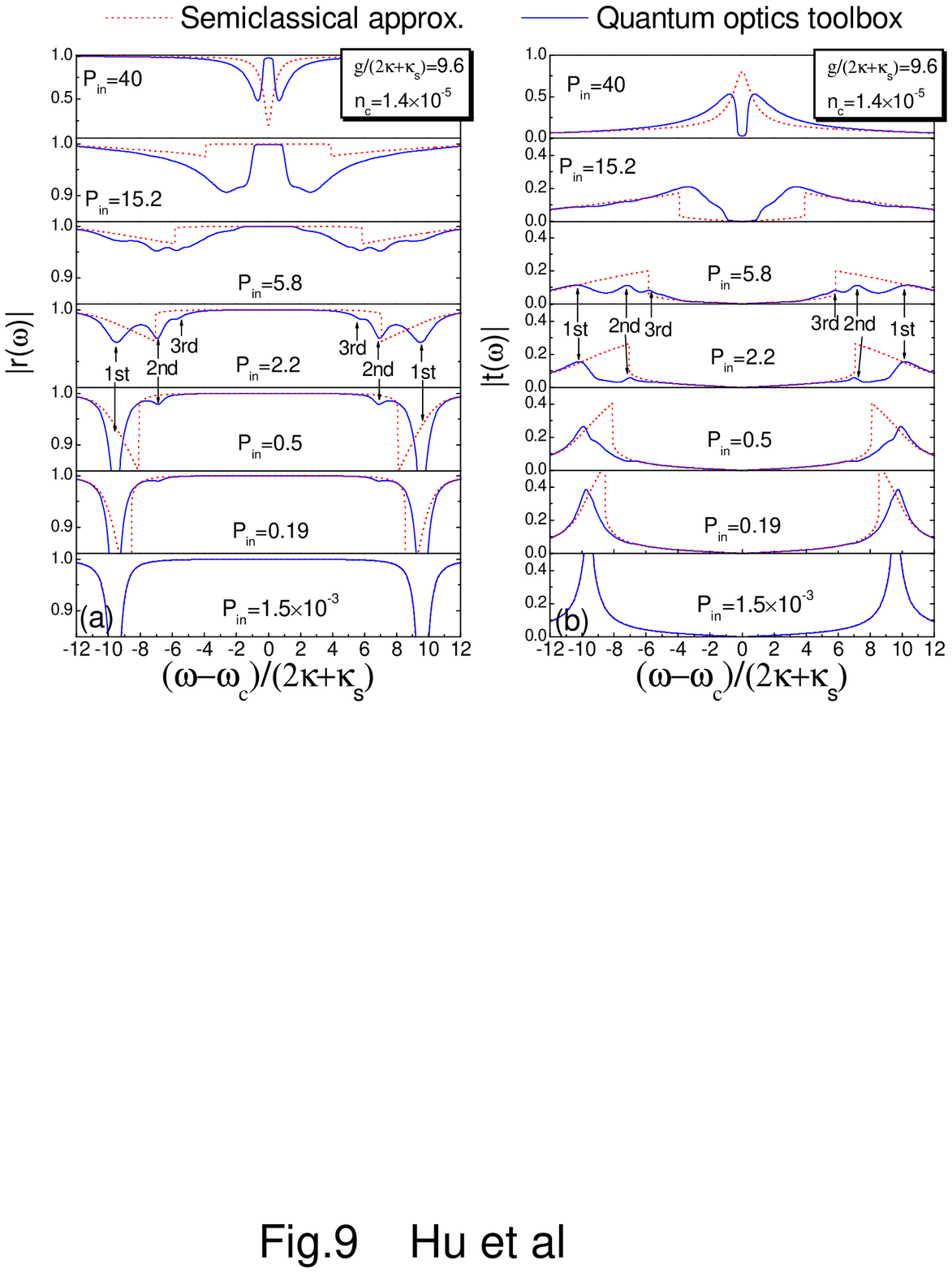}
\caption{(color online). (a) Reflectance $|r_h(\omega)|$ spectra and (b) transmittance
$|t_h(\omega)|$ spectra from a hot cavity in the strong coupling regime with
$\mathrm{g}=9.6(2\kappa+\kappa_s)$ at different input field powers.
The input power $P_{in}$
is normalized by $\kappa+\kappa_s$ (i.e., in photons per cavity lifetime).
The 1st, 2nd and 3rd manifold of dressed states are observed in certain power range.
Red dotted curves
are calculated by using Eq. (\ref{eqd2}) in semiclassical approximation, and
blue solid curves are calculated by the quantum optics toolbox.} \label{fig9}
\end{figure}

As all the dressed state resonances are situated around the edges of the non-saturation
window, the linear phase shift or the linear GFR around the cavity mode resonance are not affected
by these dressed state resonances [see Fig. 8(b)] and persist up to $P_{in}\simeq 20$
in accordance with calculations.\cite{exp4}
We notice that these dressed state resonances are observed only in a limited power range.
The input field should be strong enough to inject enough photons into cavity so that
the occupation of $|n, G\rangle$ and $|n-1, E\rangle$ states can develop.
Meanwhile, the input light should not be too strong to saturate the n-photon
transitions $|0, G\rangle \rightarrow |n, \pm\rangle$.
The higher the order of the dressed states, the smaller
this power range and the more difficult to observe the dressed state resonances.
Within this power range, the dressed state resonances
remain in the same energy position. However,
they become broader and get saturated finally, and after that all of them  merge to
two broad resonances which shift towards the cavity resonance with increasing input power.
This can be explained by the QD saturation
which reduces the coupling strength $\mathrm{g}$
to $\mathrm{g}_{eff}=\mathrm{g}\sqrt{|\langle \sigma_z\rangle|}$ as discussed in Sec.II.

Fig. 9 shows the reflection and transmission spectra at different input powers for a double-sided spin-cavity
system with $\mathrm{g}/(2\kappa+\kappa_s)=9.6$. The $n=1,2,3$ dressed state resonances are again identified
in the reflection and transmission spectra calculated by the toolbox.  The reflection and transmission
around the cavity resonance, i.e., the linear GCB are not affected by the higher-order
dressed state resonances with increasing the power up to
$P_{in}\simeq40$ in accordance with calculations. \cite{exp4}

Based on the discussions above, we can conclude that the linear GFR and GCB are
robust against the multi-photon transitions (besides the QD saturation) up to a high power where the
non-saturation window is closed.\cite{exp4}
It is quite tricky to observe these high-order dressed state resonances.
Besides the requirements of strong coupling and the resonance condition $n\omega=\omega_{n,\pm}$, lower saturation, higher coupling
strength and the right power range also need to be taken into account.
Detailed discussions on the criteria to observe these multi-photon transitions
are lengthy and go beyond the scope of this work, and will be published elsewhere.\cite{hu15}

\section{Conclusions}

We have studied the saturation nonlinear effects in QD-spin coupled cavity QED systems
using an analytical approach in the semiclassical approximation compared with a numerical
approach using Tan's quantum optics toolbox. We find that the semiclassical approximation
can be used not only in the low-power regime ($P_{in}\ll1$) and high-power regime ($P_{in}\gg1$),
but also at intermediate powers ($P_{in}\sim 1$) in a non-saturation window between
the two dressed-state resonances. In the low-power regime where
the QD is in the ground  state and the saturation is negligibly small, the semiclassical
approximation is equivalent to the weak-excitation approximation where the GFR and GCB
are linear effects across the whole frequency range.

The dressed state resonances saturate when the incident
field contains much less than one photon per cavity lifetime leading to saturation
nonlinearity in the associated GFR and GCB.
Between the dressed state resonances (i.e., within the non-saturation
window) the saturation occurs at much higher incident photon rate ( at a level of one photon
per cavity lifetime) and we can see  power independent,linear GFR and GCB
around the cavity resonance in strong coupling regime, which
are robust against the QD saturation and multi-photon transitions.
The higher the coupling strength g, the higher powers
the linear effects can retain up to.\cite{exp4}
In the Purcell regime there is no dressed state splitting
thus no non-saturation window and no linear effects exist.
We conclude that the quantum gates \cite{hu08, hu09} built from the linear effects
either in the strong coupling regime or in the low power limit are robust against
the input field intensity fluctuations, and can be safely applied for high-speed
quantum and classical information processing with varying photon rates.

The fact that there is a lower nonlinear threshold for the dressed states suggests
one could modify a relatively high power on-resonance beam using a lower power beam
resonant with the dressed states.  We are studying this \textquotedblleft transistor\textquotedblright
action and will present it in a separate paper.

\section*{ACKNOWLEDGMENTS}
We acknowledge stimulating discussions with J. Adcock and T.D. Galley.
This work is funded by ERC advanced grant QUOWSS (No. 247462) and ERANET/EPSRC project SSQN.


\begin{references}
\bibitem{exp2}The QD-cavity coupling strength is given by $\text{g}=\sqrt{e^2f/(4\epsilon_r\epsilon_0 m_0V_{eff})}$
where $f$ is the $X^-$ oscillator strength, and $V_{eff}$ is the cavity mode volume.
\bibitem{note1}The total QD dipole decay rate defined as $\gamma/2$ in our work is
equivalent to $\gamma_{\perp}$ defined in some other literatures, i.e.,
$\gamma/2=\gamma_{\parallel}/2+\gamma^* =\gamma_{\perp}$.
\bibitem{note2}The saturation nonlinearity starts when the average cavity photon
number $\langle n\rangle$ becomes larger than the critical photon number $n_c$. However,
$\langle n\rangle$ is dependent on the input power, the frequency $\omega$
and the coupling strength g, therefore for convenience we use the power $P_{in}$ that
is normalized by photons per cavity lifetime to describe the threshold of saturation
nonlinearities in our work. It is a complicated function between $\langle n\rangle$
and $P_{in}$ as described by Eqs. (\ref{eqs3b}) and (\ref{eqd3}).
\bibitem{exp1}In the strong coupling regime, $\langle \sigma_z\rangle$ and $\langle n\rangle$ can have
three solutions at some frequencies at intermediate powers. However, we take
only one of them in this work. Whether the triple stable phenomenon is real or just caused
by the semi-classical approximation is still under investigation.
\bibitem{exp3}The power dependence of the reflection or transmission spectra can be utilized
as an optical detector to measure the efficiency of input light coupling to the QD-cavity systems in experiments.
\bibitem{exp4}From Eq. (\ref{eqs3a}), Eq. (\ref{eqs3b}) and Eq.(\ref{eqd3}),
we can estimate that the non-saturation window is closed roughly
at $P_{in}=\mathrm{g}^2\gamma_{\parallel}/8\kappa \gamma(\kappa+\kappa_s)$ for single-sided cavity
structures, or at $P_{in}=\mathrm{g}^2\gamma_{\parallel}/8\kappa \gamma(2\kappa+\kappa_s)$ for
double-sided cavity structures. The higher the coupling strength g, the higher powers the
linear GFR and GCB can retain to.
\end{references}
\end{document}